\PassOptionsToPackage{dvipsnames,table,x11names}{xcolor}
\documentclass[preprint]{vgtc}
\usepackage{hyperref}

\graphicspath{{figures/}{pictures/}{images/}{./}} 

\usepackage{times}                     

\usepackage{tabularray}
\UseTblrLibrary{booktabs}

\usepackage{mathptmx}                  

\usepackage{etoolbox}

\newtoggle{engver}
\toggletrue{engver}

\usepackage{comment,amsmath,multirow,siunitx,amssymb}
\usepackage{enumitem} 
\usepackage{tabularx}
\usepackage[whole]{bxcjkjatype}

\usepackage{balance}

\usepackage{xspace}
\newcommand{\etal}{et al.\@\xspace} 
\newcommand{\eg}{e.g.\@\xspace} 

\usepackage{pdfpages}
\usepackage{pgffor}
\usepackage{url}
\pdfximage{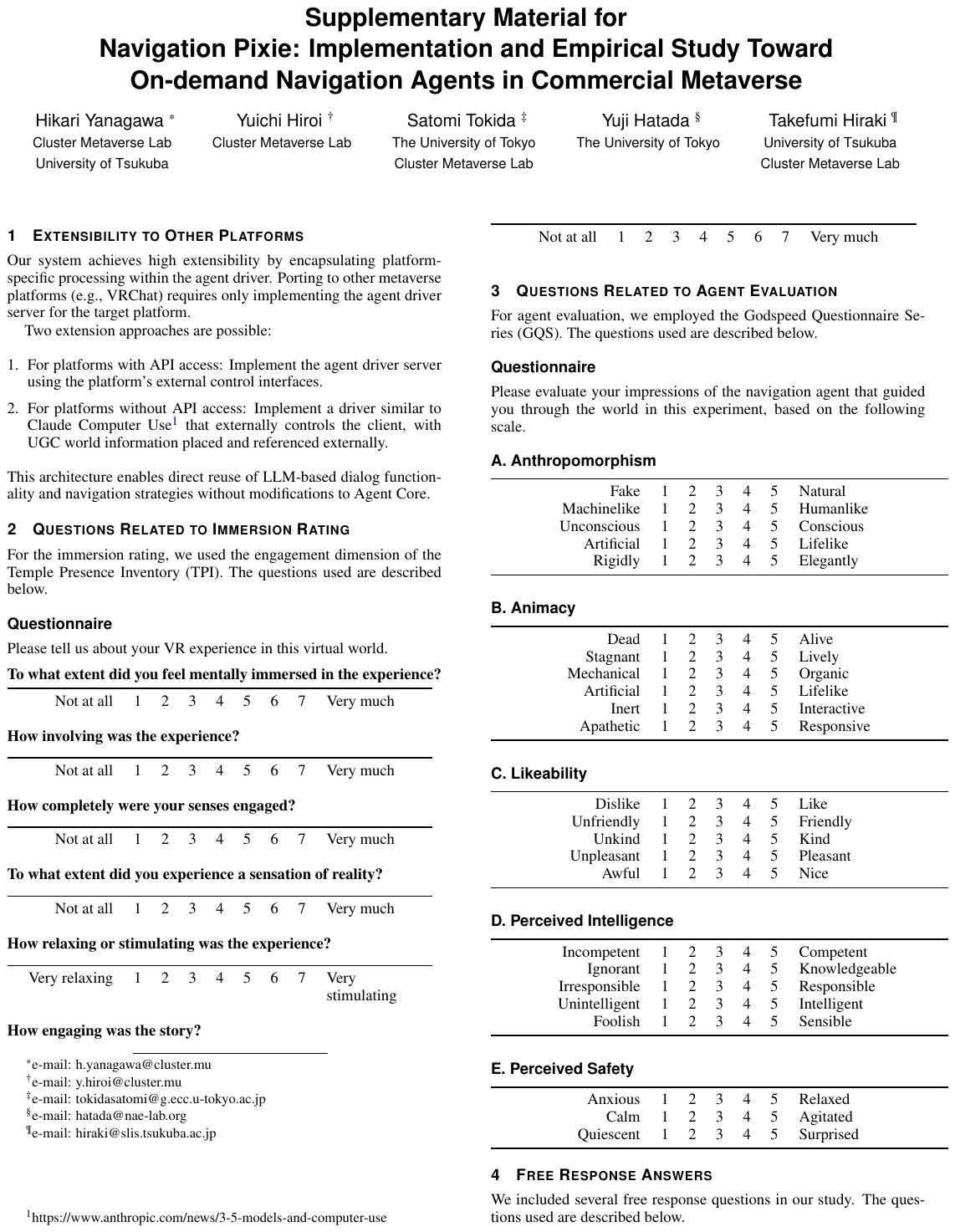}
\def\numbersupplementpages{\the\pdflastximagepages}
\newif\ifarXiv
\arXivtrue

\onlineid{1789}

\vgtccategory{Research}

\title{Navigation Pixie: Implementation and Empirical Study Toward On-demand Navigation Agents in Commercial Metaverse}

\author{
Hikari Yanagawa \thanks{e-mail: h.yanagawa@cluster.mu}\\ \scriptsize Cluster Metaverse Lab \\ \scriptsize University of Tsukuba%
\and Yuichi Hiroi \thanks{e-mail: y.hiroi@cluster.mu}\\ \scriptsize Cluster Metaverse Lab
\and Satomi Tokida \thanks{e-mail: tokidasatomi@g.ecc.u-tokyo.ac.jp}\\ \scriptsize The University of Tokyo \\ \scriptsize Cluster Metaverse Lab %
\and Yuji Hatada \thanks{e-mail: hatada@nae-lab.org}\\ \scriptsize The University of Tokyo %
\and Takefumi Hiraki \thanks{e-mail: hiraki@slis.tsukuba.ac.jp}\\ \scriptsize University of Tsukuba \\ \scriptsize Cluster Metaverse Lab
}

\teaser{
  \centering
  \includegraphics[width=0.93\linewidth]{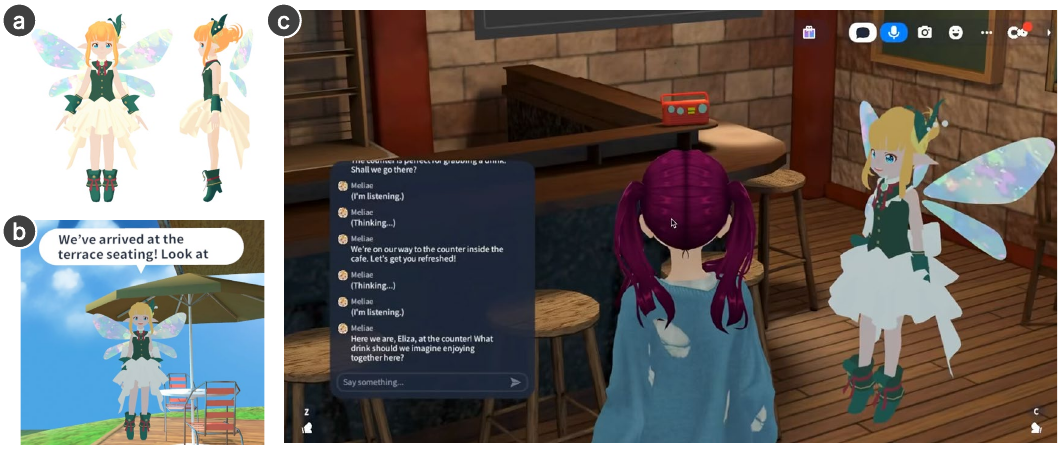}
  \caption{Our on-demand navigation agent implemented on a commercial metaverse platform. (a) Front and side views of the agent's 3D avatar design. The pixie-like design balances non-human, fictional characteristics with a non-verbal communication function for a tour guide. (b) The agent explaining the patio seating experience through voice and text, demonstrating spatial awareness and providing contextual information through natural language. (c) Actual interaction between a user and the agent on the commercial metaverse platform, showing the agent guiding the user to the bar and suggesting beverage options.
}
  \label{fig:teaser}
}

\abstract{
While commercial metaverse platforms offer diverse user-generated content, they lack effective navigation assistance that can dynamically adapt to users' interests and intentions. Although previous research has investigated on-demand agents in controlled environments, implementation in commercial settings with diverse world configurations and platform constraints remains challenging.

We present Navigation Pixie, an on-demand navigation agent employing a loosely coupled architecture that integrates structured spatial metadata with LLM-based natural language processing while minimizing platform dependencies, which enables experiments on the extensive user base of commercial metaverse platforms. Our cross-platform experiments on commercial metaverse platform Cluster with 99 PC client and 94 VR-HMD participants demonstrated that Navigation Pixie significantly increased dwell time and free exploration compared to fixed-route and no-agent conditions across both platforms. Subjective evaluations revealed consistent on-demand preferences in PC environments versus context-dependent social perception advantages in VR-HMD.
This research contributes to advancing VR interaction design through conversational spatial navigation agents, establishes cross-platform evaluation methodologies revealing environment-dependent effectiveness, and demonstrates empirical experimentation frameworks for commercial metaverse platforms.
}

\keywords{On-demand Navigation Agent, Commercial Metaverse Platforms, User-generated Content}

\begin{document}



\firstsection{Introduction}\label{sec:intro}
\maketitle
The evolution of commercial metaverse platforms such as VRChat\footnote{https://hello.vrchat.com/}, Resonite\footnote{https://store.steampowered.com/app/2519830/Resonite/}, and Cluster\footnote{https://cluster.mu/en} has facilitated the formation of diverse virtual spaces centered on user-generated content (UGC)~\cite{Ritterbusch2023-ct}. These UGC-based worlds encompass a wide range of purposes and structures, from art exhibitions to social events and gamified experiences, that encourage free-roaming and exploration within metaverse environments. However, this diversity presents challenges for visitors, who often struggle to comprehend destinations and construct meaningful exploration experiences.

In such environments, there is a growing demand for on-demand navigational assistance agents that can instantly respond to user interests and situations. Traditional agent functionality in commercial metaverse platforms has been implemented primarily through ``non-intelligent agents,'' including script-based guides that follow predetermined routes and provide limited keyword responses~\cite{chiattaro2004navigating, jan2009virtual}. However, these approaches inadequately address users' ambiguous utterances or evolving interests, often limiting the exploration experience to superficial interactions.

The rapid development of Large Language Models (LLMs) has dramatically increased the feasibility of agents with real-time contextual understanding and natural dialog capabilities that were previously difficult to achieve~\cite{prpa2023ellmat, wan2024building, wang2024enhancing}. LLM-based agents that integrate speech, language, and spatial understanding can flexibly select actions based on user requests, effectively acting as ``exploration companions''~\cite{10.1145/1067860.1067867}. Such agents go beyond simply providing functionality and have the potential to transform the fundamental quality of the user experience. However, previous research on navigational assistance agents has mostly focused on specific researcher-designed systems or VR spaces. While these efforts have provided valuable insights into agent design and evaluation, most rely on specialized environment configurations that lack the portability required for direct implementation in commercial metaverse environments.

Furthermore, metaverse spaces centered on UGC present additional complexities: world configurations and interaction designs vary among creators, while platform-specific API specifications and development constraints impose further limitations. As a result, design insights for implementing agents that can be deployed by non-specialists remain limited, especially when considering these operational complexities and differences.
This situation significantly impedes the enhancement of user experience in commercial metaverse platforms where UGC constitutes the core component.

This research addresses these challenges by proposing ``Navigation Pixie,'' an on-demand navigation agent with a loosely coupled architecture that users can implement themselves on commercial metaverse platforms. By integrating structured spatial information from metaverse environments with the natural language understanding capabilities of large language models, this agent provides flexible guidance that responds to user requests while accommodating the diverse configurations of different UGC worlds.

Beyond demonstrating technical feasibility, this research examines how Pixie's intelligent flexibility impacts the actual user experience by comparing it to script-based fixed route agents (conventional ``non-intelligent agents''), measuring changes in exploration behavior and immersive experiences across both PC client and VR-HMD platforms. Notably, these experiments used fully operational implementations on a commercial metaverse platform, Cluster, with real users. This approach allowed us to gather 99 PC client participants within 2 days and 94 VR-HMD participants within 5 days, thus achieving scalable empirical evaluation, and demonstrates the potential of using commercial metaverse platforms as socially implemented research infrastructure.

This paper discusses the potential qualitative transformation of metaverse experiences through LLM-based interactive agents and the methodological developments in agent research utilizing metaverse spaces, through the design architecture of Navigation Pixie, its specific implementation structure, and analysis based on user experiments.

\section{Related Work}\label{sec:related-work}
\subsection{Embodied Navigation Agents in VR}
The development of embodied conversational agents that function as tour guides has been driven by spatial disorientation and wayfinding difficulties in large-scale 3D environments. Early implementations emerged in the form of physical museum tour robots around 2000~\cite{BURGARD19993, Trahanias2000tourbot},  while the first virtual implementations appeared with Jan~\etal interactive agents in Second Life~\cite{jan2009virtual}. The embodiment of these agents significantly improves the perception of companionship and immersion~\cite{10.1111/0022-4537.00153, 10.3389/frvir.2021.786665}. In addition, their nonverbal communication capabilities, such as gaze direction and pointing gestures, enable them to control attention and time information similar to human museum curators~\cite{chiattaro2004navigating, Kadobayashi2000seamless}.

Traditional virtual guides have predominantly employed fixed-route approaches, ensuring comprehensive exhibit coverage while constraining user exploration freedom~\cite{chiattaro2004navigating, 10.2312:egve.20191276, 10.1145/3472306.3478339}. In contrast, companion-type agents that dynamically respond to user interests offer greater flexibility, despite implementation complexity~\cite{10.3389/frvir.2023.1334795}. While systems like English-learning NPCs on VRChat demonstrate conversational capabilities~\cite{prpa2023ellmat}, they typically prioritize educational dialogue over spatial navigation functionality.

Recent advances in LLMs have accelerated the research of embodied AI agents that integrate language, vision, and spatial information~\cite{gu-etal-2022-vision, Naveed2023}, with preliminary investigations conducted in a controlled virtual museum environment~\cite{10.1145/3613904.3642235}. Our research builds on these approaches by implementing embodied agents that integrate LLMs with spatial information on commercial VR platforms.

\subsection{On-demand Navigation}
VR environments present unique challenges regarding self-localization and route formation due to their structural flexibility compared to physical spaces. To address these challenges while preserving user autonomy, on-demand approaches that provide guidance upon request have emerged. Early work introduced adaptive guides in web-based virtual museums, proposing various information delivery methods based on user models and browsing history~\cite{marucci2002design}. More recent developments include a visual guide agent by Collins~\etal for visually impaired users in social VR spaces, which provides environment descriptions and navigation support in response to user requests~\cite{10.1145/3597638.3608386}.

Ye~\etal provide positioning optimization for guide agents that facilitate natural dialogue-based guidance in VR~\cite{Ye2021paval}, while Hammady~\etal developed guides that integrate real exhibition spaces with MR in HoloLens, improving visitor satisfaction through rich 3D displays and audio guidance~\cite{Hammady2020MuseumEye}. However, architectural design and large-scale user evaluation methodologies for on-demand agents in diverse UGC environments within commercial metaverse platforms remain insufficiently explored.

\subsection{Experimentation in Metaverse Platforms}
Recently, commercial metaverse platforms such as VRChat and Cluster have attracted attention as large-scale, remote VR experimental infrastructures due to their diverse and continuously connected user bases~\cite{10.1145/3472617, 10843684}. Using commercial platforms effectively incorporates realistic diversity, which is difficult to replicate in laboratory environments~\cite{10.1145/3472306.3478339, 10.1145/3611659.3615697}. Saffo~\etal~\cite{Saffo2020CrowdsourcingVirtualReality, Saffo2021TwoPathsTowards} and Kurai~\etal~\cite{10843684, Kurai2025-bf} demonstrated the potential of crowdsourced VR experiments by collecting user experience data and behavioral logs through commercial metaverse platforms. However, they also noted that reliance on the existing functionality of commercial metaverses creates limitations in experimental world design and data collection. Additionally, platform-specific APIs and version constraints pose challenges that necessitate purpose-appropriate architectural design.

UGC-centered environments, in particular, require technical innovation for external control and data acquisition from independently developed elements~\cite{10.1145/3665318.3678230}. Although experimental platforms such as Ubiq~\cite{Friston2021-td, Steed2022, Numan2023-nx, Giunchi2024-gt} and Ouvrai~\cite{cesanek2024ouvrai} have been developed to address these challenges, they have yet to achieve user bases comparable to those of commercial systems in terms of diversity and scale. Our research implements a loosely coupled architecture that reduces dependence on platforms, creating an agent evaluation environment that leverages commercial metaverse platforms with large user bases.

\section{System Architecture}\label{sec:system-arch}
\begin{figure*}[t]
    \centering
    \includegraphics[width=1\linewidth]{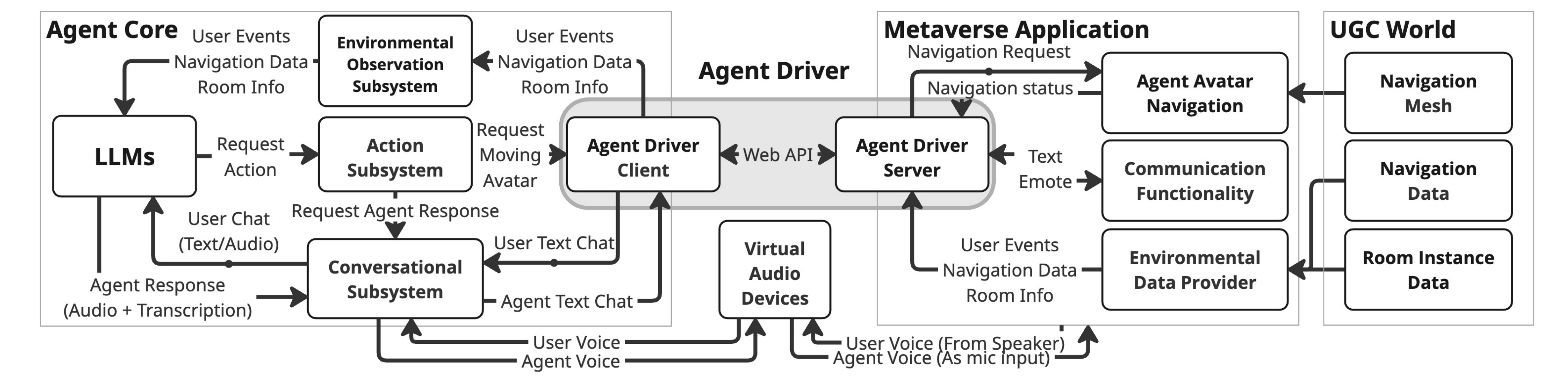}
    \vspace*{-3mm}
    \caption{Architectural diagram of the Navigation Pixie system. The commercial metaverse platform and the external Agent Core, which serves as the cognitive center of the agent, are loosely coupled only through the Agent Driver, a compatibility layer that ensures platform independence. This design achieves effective navigation support by integrating spatial understanding and natural language processing while clearly separating platform-dependent processes.}
    \label{fig:system_overviews}
    \vspace{-3mm}
\end{figure*}

This section describes the Navigation Pixie architecture, an on-demand navigation agent designed for commercial metaverse platforms, which enable enables platform-independent operation through loose coupling.

\subsection{Design Principles}
Our Navigation Pixie architecture follows four core principles:
\begin{enumerate}[leftmargin=*]
\setlength{\parskip}{0cm}
   \item \textbf{Platform Independence}: To enable deployment across multiple metaverse environment, we isolate platform-specific code in dedicated modules and establish standardized interfaces.
   
   \item \textbf{UGC Compatibility}: To accommodate dynamically changing UGC, we provide platform-independent tools that allow creators to embed structured spatial information (\eg locations and detailed descriptions) into their worlds.
   
   \item \textbf{LLM Adaptability}: To adapt rapid model evolution of LLMs, we implement core functionality through rapid prompt engineering rather than custom model training.
   
   \item \textbf{Operational Sustainability}: To ensure compatibility with frequently updated platforms, we define common interfaces to essential platform functionality.
\end{enumerate}

\begin{figure}[t]
    \centering
    \includegraphics[width=1\linewidth]{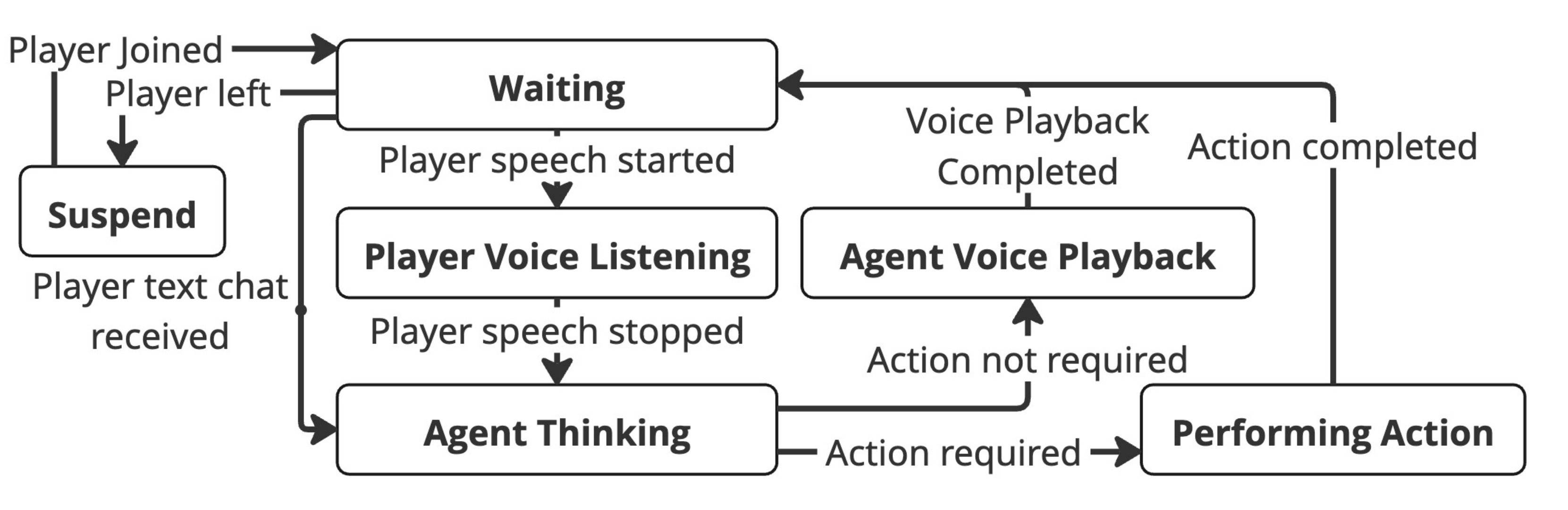}
    \vspace*{-3mm}
    \caption{State transition diagram of the Navigation Pixie agent. Agent operation is managed through transitions between six primary states (suspend, waiting, player voice listening, agent thinking, agent voice playback, and performing action), with state changes governed by user input (voice and text), agent processing status, and action requirements.}
    \label{fig:avatar_states}
    \vspace{-5mm}
\end{figure}

\subsection{Architectural Overview}
As shown in Figure~\ref{fig:system_overviews}, Navigation Pixie consists of four primary components that communicate through the standardized agent driver interface. This architecture handles the complexities of LLM integration and cross-platform compatibility, allowing UGC creators to integrate navigation agents by embedding spatial metadata with less technical expertise.

\paragraph{Agent Driver} forms the compatibility layer between Agent Core and metaverse platforms, serving as an abstraction layer that enables platform independence. This interface provides four core functions:  \textbf{(i) Environmental Data Capture} retrieves and normalizes structured spatial information, room metadata, and user access information from the metaverse platform. \textbf{(ii) Event Propagation} manages real-time events such as user entry/exit notifications and text communications via publish-subscribe patterns. \textbf{(iii) Avatar Control} provides abstract operations for position control, orientation management, destination setting, and path status monitoring while hiding platform-specific implementation details. \textbf{(iv) Communication Access} exposes standardized methods for text messaging and emote execution.

Agent Driver processes two types of structured spatial data essential for spatial understanding: \textbf{(i) Navigation Mesh} represents traversable areas for path computation, and \textbf{(ii) Navigation Data} provides structured information including spatial names, location identifiers, coordinates, and semantic descriptions for LLM context awareness.

\paragraph{Agent Core} serves as the cognitive center of the navigation system, comprising five integrated subsystems: \textbf{(i) Agent Driver Client} communicates with the server through WebAPI, sending action commands and receiving environmental data. \textbf{(ii) Environmental Observation Subsystem} receives user events, navigation data, and room information, providing them to the LLM. \textbf{(iii) LLM} makes decisions based on environmental data and user input, issuing responses and action instructions. \textbf{(iv) Action Subsystem} generates specific action instructions based on LLM decisions. \textbf{(v) Conversational Subsystem} manages user input and agent dialog while generating responses.

\paragraph{Metaverse Application} implements platform-specific functionality by encapsulating proprietary platform APIs within standardized agent driver interfaces. This component comprises four integrated subsystems: \textbf{(i) Agent Driver Server} processes requests from the agent driver client and handles conversion to platform-specific APIs. \textbf{(ii) Avatar Navigation} controls avatar movement to specified locations and provides navigation feasibility and remaining distance. \textbf{(iii) Communication Functionality} manages standard user communication features including text chat and emotes. \textbf{(iv) Environmental Data Provider} collects spatial information from UGC World and provides it to the Agent Driver Server.

\paragraph{UGC World} contains spatial information created by users using the game engine or platform features, including navigation meshes that define traversable areas, navigation data with coordinates and semantic descriptions, and room instance metadata such as room names.

\paragraph{Virtual Audio Devices} mediate voice communication between users and the agent, enabling natural conversational interaction.

\section{Implementation}
We implemented the architecture described in Sec.~\ref{sec:system-arch} on the commercial metaverse platform Cluster.

\begin{figure*}[t]
    \centering
    \includegraphics[width=1\linewidth]{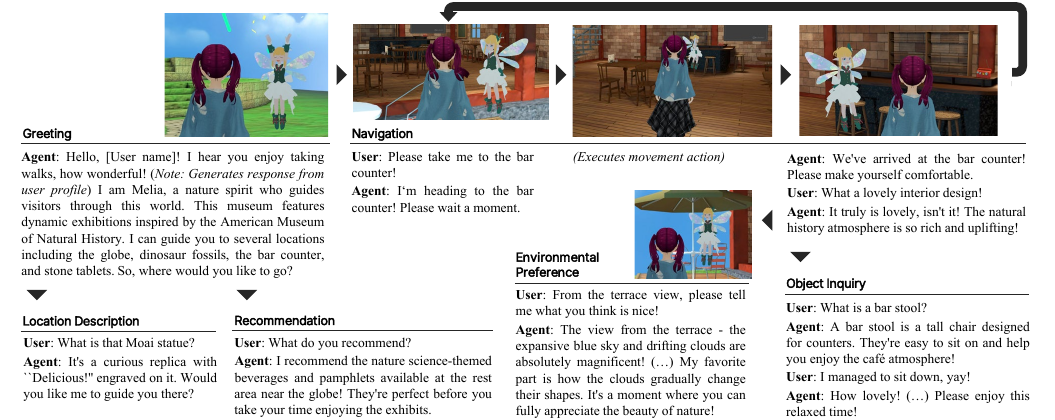}
    \vspace*{0.5mm}
    \caption{User experience flow and selected examples of conversations between Navigation Pixie (agent) and users. Through various dialogue patterns, the agent demonstrates capabilities such as using user profile information, performing spatial movements, providing detailed explanations about the environment, expressing personal views, and maintaining natural conversation.}
    \vspace*{-2mm}
    \label{fig:conversation_examples}
\end{figure*}

\subsection{Component Implementation}
Agent Core was implemented in Python and deployed on AWS EC2, using a two-tier LLM architecture: OpenAI GPT-4o Realtime (gpt-4o-realtime-preview-2024-12-17) for speech input/output processing as primary, and GPT-4o (gpt-4o-2024-11-20) for decision-making as secondary. 
These models integrate through OpenAI's function calling API, enabling both speech interaction and sophisticated decision processes. 

Environmental Observation System formats spatial information in custom JSON structures, while Agent Driver Client communicates via REST API and WebSocket for action commands and environmental data exchange.

Metaverse Application was implemented by customizing the client source code provided with permission from Cluster, Inc., the developer of Cluster. We extended this Unity-based client with the agent driver server functionality, implementing text chat sending/receiving, avatar movement control, environmental data collection/transmission, and user event notification as Web API services.

Avatar Navigation utilizes Unity's NavMesh agent for movement control, while Communication functionality adapts Cluster's text chat and emote features with the addition of user attention-switching capabilities based on avatar state.
Spatial information retrieval leverages Cluster's world loading mechanism, referencing Unity components and server data for room and user information.

Note that our implementation achieves high extensibility by encapsulating platform-specific processing within the agent driver. Porting to other metaverse platforms (\eg, VRChat) requires only implementing Agent Driver Server for the target platform. Detailed extension procedures for other platforms are provided in Supplementary Material.

\subsection{Avatar Design}
We selected a pixie-like avatar design to simplify agent driver implementation (Fig.~\ref{fig:teaser}). through gliding movement, eliminating walk/jump animation requirements. The design prioritizes approachability with intelligence signaling, visual differentiation from human avatars, high visibility, and non-verbal expression capabilities suitable for natural language interaction.

\subsection{State Management and Agent Behavior Control}
Figure~\ref{fig:avatar_states} shows the state transition of the agent. Agent operation follows six state transitions: \textbf{(i) Suspend} waits for user entry, \textbf{(ii) Waiting} displays ``(Please speak)'' to encourage interaction, \textbf{(iii) Player Voice Listening} detects and processes user's voice input, \textbf{(iv) Agent Thinking} processes utterances with LLMs while displaying ``(Thinking)'', \textbf{(v) Agent Voice Playback} plays LLM responses until completion, and \textbf{(vi) Action Execution} performs movements before returning to waiting state.

\subsection{Agent Performance Evaluation}
\paragraph{Response Time}
We measured the response time of the agent by measuring duration from user's voice utterance completion to response playback initiation. Based on 60 samples, the average response time was 6.40~$\pm$~3.61 seconds, including speech-to-text transcription, response generation, and speech synthesis. This result is longer than turn-taking intervals in natural conversation (0.5-1 second)~\cite{sacks1974simple}. However, this latency reflects GPT-4o Realtime processing capabilities and can improve with future model evolution. To mitigate the perceived delay in the dialog, we implemented filler messages (``Thinking...'') and head tilt animations~\cite{1360017285628554240}.

\paragraph{Agent Conversation Examples}
Figure~\ref{fig:conversation_examples} demonstrates spatial guidance capabilities, environmental query responses, and the natural conversational flow of our agent through representative user interactions.

\section{User Study}
This section details our user study of Navigation Pixie, addressing the following research questions:

\begin{description}[leftmargin=*]
\item[RQ1] To what extent do on-demand navigation agents improve exploration behavior and dwell time in VR environments, including metaverse platforms?
\item[RQ2] What design elements, such as interaction format, responsiveness, and navigation style, most effectively increase user interest and immersion in VR environments?
\end{description}

\subsection{Experiment Setup}
\subsubsection{Experimental Design}
We conducted two separate experiments to evaluate platform-specific effects: a PC client study and a VR-HMD study. Participants were randomly assigned to experimental conditions: 
\begin{enumerate}[label=(\Alph*), leftmargin=*] 
\setlength{\parskip}{0cm}
\item \textbf{On-Demand Agent}: Our Navigation Pixie implementation integrated natural language dialog with spatial navigation capabilities. Users freely conversed with the agent and requested guidance to specific destinations, with the agent responding using LLM contextual understanding and spatial structure information.
\item \textbf{Fixed-Route Agent}: Guided users along predetermined routes, moving to destinations upon user instructions (\eg, ``OK'') and providing explanations at each location. While using the same spatially informed LLM as Group A, the route remained fixed. After completing planned exploration, users could explore freely.
\item \textbf{Control}: Users entered and explored the room without any agent or guidance. This condition was only included in the PC client study.
\end{enumerate}
This design enables comparison of agent presence (A \& B vs. C) and dialog capability (A vs. B) on user experience.

\subsubsection{Experimental Environment}
\begin{figure}[t]
    \centering
    \includegraphics[width=1\linewidth]{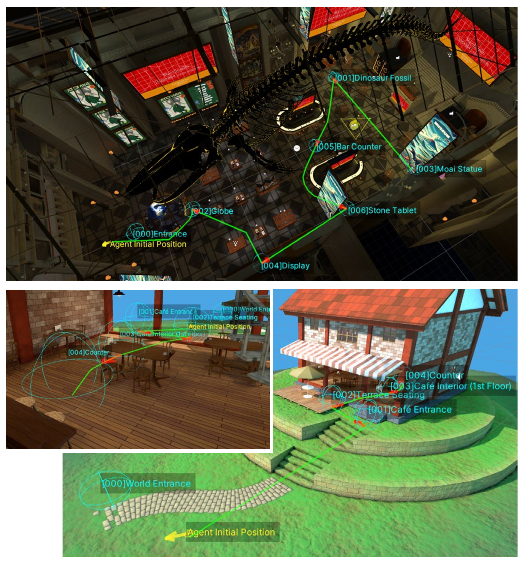}
    \vspace*{-3mm}
    \caption{Two virtual environments and navigation points used in the experiment. (Top) The \textbf{Museum} environment has 7 navigation points centered around exhibits such as dinosaur fossils and globes. (Bottom) The \textbf{Ruina} sky cafe environment includes 5 navigation points across a social space with terrace seating and counters. Both environments provide structured spatial information and semantic descriptions to the LLM, enabling navigation from the agent's starting position to each point. Green lines indicate the predetermined route for group B (fixed route agent).}
    \label{fig:exp_world_info}
    \vspace{-3mm}
\end{figure}

The experiment used two UGC worlds in Cluster (Fig.~\ref{fig:exp_world_info}): \textbf{Museum} featured an exhibition gallery (36.1~m $\times$ 66.0~m) with 7 navigation points, while the world named \textbf{Ruina}\footnote{https://cluster.mu/w/1aaa3e6e-4823-4339-ab51-ca37cd461902} offered a sky cafe (37.0~m $\times$ 51.7~m) with 5 navigation points. Both environments provided comparable explorable areas with contrasting contexts (indoor/cultural vs. outdoor/social) and structured spatial information for LLM initialization. Participants were able to communicate with agents using either text chat or voice.

\subsubsection{Participants}
Participants were recruited through the Cluster's official events and social media channels. The PC client study (1000 JPY compensation) recruited 135 interested participants over 2 days, with 99 completing the four-day experimental period. The VR-HMD study provided higher compensation (2000 JPY) due to limited HMD availability, recruiting 132 participants over 5 days and yielding 94 valid datasets. Both studies applied identical exclusion criteria for no-shows, technical problems, and incomplete surveys.

\begin{figure}[t]
\centering
\includegraphics[width=1\linewidth]{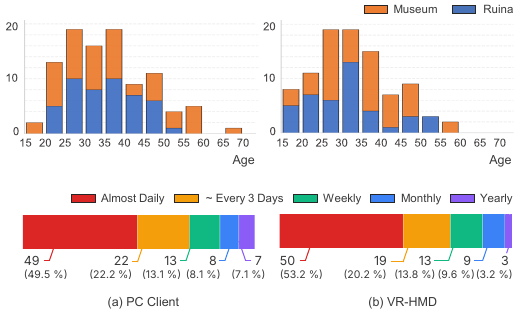}
\vspace*{-3mm}
\caption{Participant demographics for (a) PC client study and (b) VR-HMD study. Top panels show age distribution by environment (Museum in orange, Ruina in blue). Bottom panels show frequency of logins to metaverse platforms with participant counts and percentages.}
\label{fig:participants_info}
\vspace{-3mm}
\end{figure}

Fig.~\ref{fig:participants_info} shows the demographics of both studies. The PC client participants (age: 36.29~$\pm$~12.01 years, range 16-65; gender: 59 males, 36 females, 4 others) and VR-HMD participants (age: 32.6~$\pm$~9.89 years, range 16-59; gender: 55 males, 23 females, 16 others) demonstrated similar demographic profiles. Participant distribution across conditions was: PC client study - Museum (A:18, B:17, C:17) and Ruina (A:15, B:15, C:17); VR-HMD study - Museum (A:26, B:26) and Ruina (A:21, B:21). Most participants were regular users of the metaverse platform and were familiar with virtual environments.

\subsubsection{Experimental Protocol}
The experiment followed three phases for both studies:
\begin{enumerate}[leftmargin=*] 
\setlength{\parskip}{0cm}
\item \textbf{Pre-participation Preparation}: Participants reviewed the experimental overview, completed a consent form on a reservation page, and reserved a time slot. Upon completion of the reservation process, participants were given URLs for randomly assigned world types and conditions.
\item \textbf{Experiment}: Participants entered worlds via distributed URLs at reserved times. Groups A and B encountered agents that provided condition-specific greetings and instructions upon entry. Although participants can leave the world at any time, all participants were asked to leave after 30 minutes via chat.
\item \textbf{Experiment Completion}: Upon exit, participants were redirected to a Google Form to provide demographic information and subjective ratings.
\end{enumerate}

\begin{figure*}[t]
    \centering
    \includegraphics[width=1\linewidth]{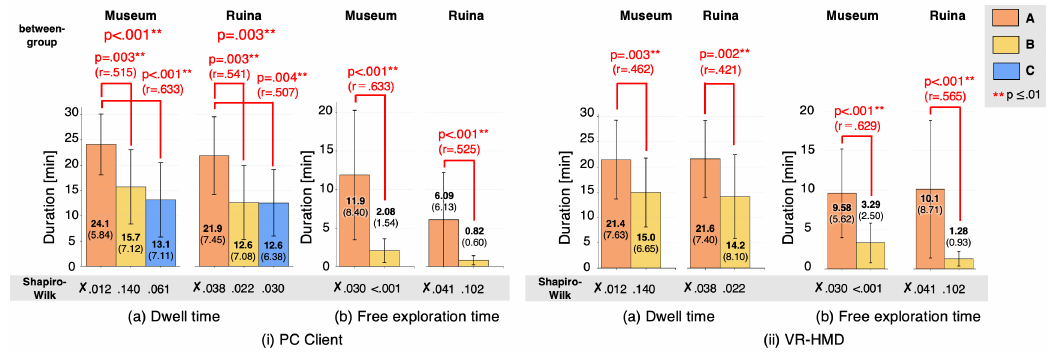}
    \caption{Behavioral metrics across experimental conditions and environments for (i) PC Client and (ii) VR-HMD studies: (a) dwell time and (b) free exploration time. Statistical significance and effect sizes shown above brackets. The legend for each statistic represents the group's name (A: On-Demand Agent, B: Fixed-Route Agent, C: Control)}
    \label{fig:exp_explore_time}
    \vspace*{-2mm}
\end{figure*}

\subsubsection{Evaluation Metrics}
Metrics evaluated navigation effectiveness corresponding to each RQ:
\paragraph{RQ1 (Exploration Behavior) Metrics}
\begin{itemize}[leftmargin=*]
\setlength{\parskip}{0cm}
\item \textbf{Dwell Time}: Duration from user entry to exit.
\item \textbf{Free Exploration Time}: Cumulative duration from when the agent completed movement to a destination and finished speaking until the user either requested another navigation instruction or exited. This metric was analyzed to eliminate the possibility that longer dwell times merely resulted from extended LLM agent speech duration. 
\end{itemize}
\paragraph{RQ2 (Immersion and Agent Evaluation) Metrics}
\begin{itemize}[leftmargin=*]
\setlength{\parskip}{0cm}
\item \textbf{Immersion Rating}: Using items from engagement dimension of the Temple Presence Inventory (TPI)~\cite{lombard2009measuring}, which measures how deeply users engage and focus during media experiences. We evaluated immersion, involvement, sensory engagement, reality sensation, relaxation/stimulation, and story appeal using 7-point Likert scales (see Supplementary Material Sec.~2).
\item \textbf{Agent Evaluation}: Using Godspeed Questionnaire Series (GQS)~\cite{bartneck2023godspeed} to evaluate impressions of the agents across five dimensions: anthropomorphism, animacy, likeability, perceived intelligence, and perceived safety, assessed on 5-point Likert scales (see Supplementary Material Sec.~3).
\item \textbf{Free Response Answers}: Open-ended questions regarding exploration motivation (see Supplementary Material Sec.~4).
\end{itemize}

\subsubsection{Analysis Methods}
Statistical analyses employed appropriate tests for between-subjects comparisons with significance set at $p < .05$. We first assessed normality using Shapiro-Wilk tests and variance homogeneity using Bartlett's (three-group) or F-tests (two-group).
For three-group test, when normality and variance homogeneity were confirmed, we used one-way ANOVA with Tukey HSD post-hoc tests; otherwise, Kruskal-Wallis tests with Holm-corrected Wilcoxon rank-sum post-hoc comparisons.
For two-group test, we applied Student's t-test when both assumptions were met, Welch's t-test when only normality held, or Wilcoxon rank-sum test when normality was violated.

\subsection{Results}
\begin{figure}[t]
    \centering
    \includegraphics[width=1\linewidth]{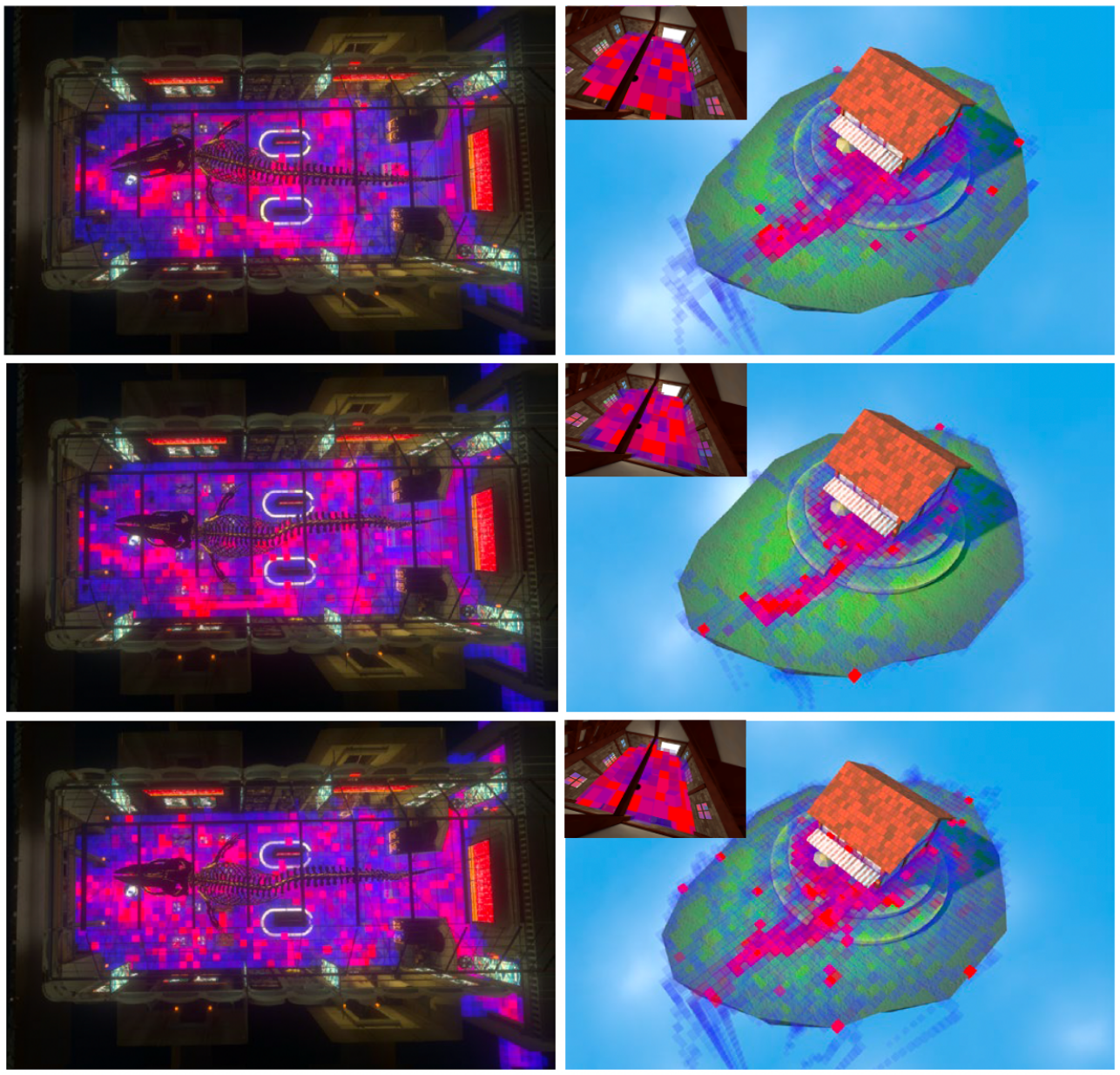}
    \vspace*{-3mm}
    \caption{Heatmaps of user dwell distribution of user trajectories across experimental conditions in PC Client: (top) Group A, (middle), B, and (bottom) C, visualizing exploration patterns in (left) Museum and (right) Ruina. Red areas show long stay locations while blue areas indicate brief visits.}
    \label{fig:exp_heatmaps}
    \vspace{-3mm}
\end{figure}

\subsubsection{Cross-Platform Behavioral and Perceptual Results}
\paragraph{Dwell Time} From Fig.~\ref{fig:exp_explore_time} (a), both platforms demonstrated consistent agent effectiveness in extending user engagement. Group A significantly extended dwell time compared to Groups B and C in PC environments, with VR-HMD results confirming this pattern between Groups A and B. 

\paragraph{Spatial Exploration Patterns} Fig.~\ref{fig:exp_heatmaps} visualizes the distribution of user dwell times as heat maps in PC Client setup, revealing distinct exploration patterns: Group B concentrated along predetermined paths, Group C exhibited limited exploration in restricted areas, while Group A demonstrated extensive spatial coverage with prolonged engagement at key interest points.

To quantify spatial utilization patterns, we calculated spatial entropy of the dwell time distributions of the PC client users. Entropy was calculated using the formula $H = -\sum p_i \log p_i$, where $p_i$ represents the proportion of dwell time spent in each spatial area. Higher entropy values indicate distributed spatial utilization, while lower values suggest concentrated exploration. Since agent speech constrains users to the same spatial region, we calculated spatial entropy based on pure movement and observation time, excluding agent speech duration. 
The results revealed the following entropy patterns of C (6.49) > B (5.82) > A (5.75) in Museum and C (5.57) > B (5.30) > A (4.90) in Ruina.

These results confirm that Group C exhibited dispersed, undirected exploration behavior as expected from the heatmap visualization. However, Group A recorded the lowest entropy values, despite demonstrating multiple circulation routes in the heatmaps. This suggests that users engage in deeper observation at specific locations after talking to the on-demand agent, which indicates that conversational interaction promotes focused spatial engagement beyond simple navigation guidance.




\paragraph{Free Exploration Time} From Fig.~\ref{fig:exp_explore_time} (b), self-directed exploration behavior showed remarkable consistency across platforms. Group A participants engaged in substantially longer periods of autonomous exploration after agent guidance completion compared to Group B participants in both PC and VR-HMD conditions, with both environments demonstrating this effect across platforms.

\begin{figure*}[t]
    \centering
    \includegraphics[width=1\linewidth]{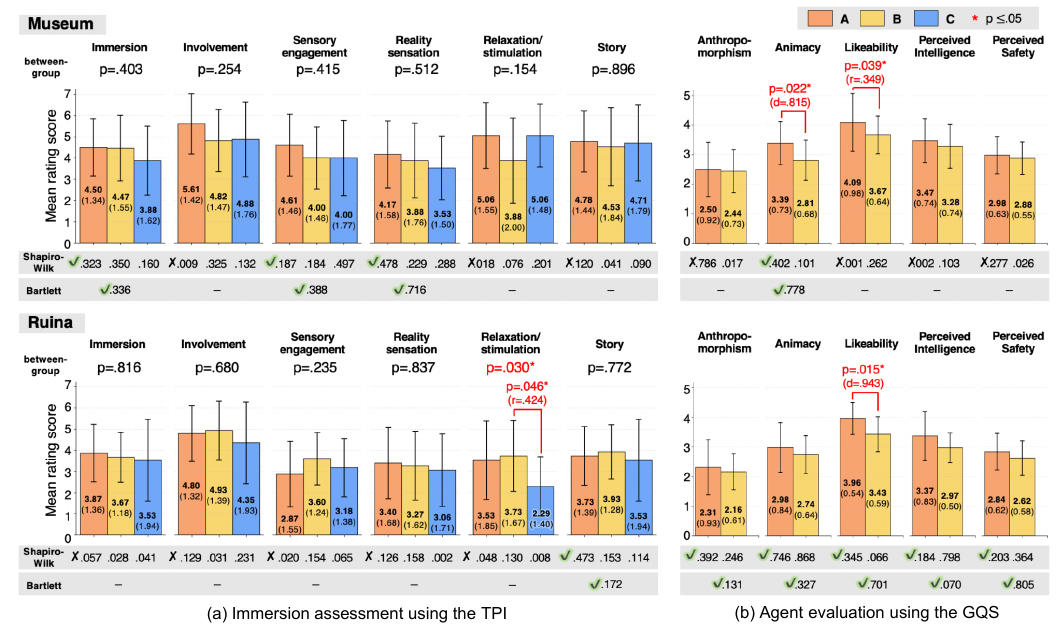}
    \vspace*{-3mm}
    \caption{Comparison of immersion experience and agent ratings in Museum (top) and Ruina environments (bottom) in PC client study. Bar graphs show means with standard deviations in parentheses and corresponding error bars. Statistical test selection was based on Shapiro-Wilk normality tests and Bartlett's tests for variance homogeneity, with results displayed at the bottom of each panel. Significant differences between groups are indicated by lines with red asterisks (*) based on $p \leq .05$. (a) Immersion assessment using the Temple Presence Inventory (TPI), with p-values below each dimension representing between-group difference test results. (b) Agent evaluation using the Godspeed Questionnaire Series (GQS).}
    \label{fig:questionnaire_results_pc}
    \vspace{-3mm}
\end{figure*}

\begin{figure*}[t]
    \centering
    \includegraphics[width=1\linewidth]{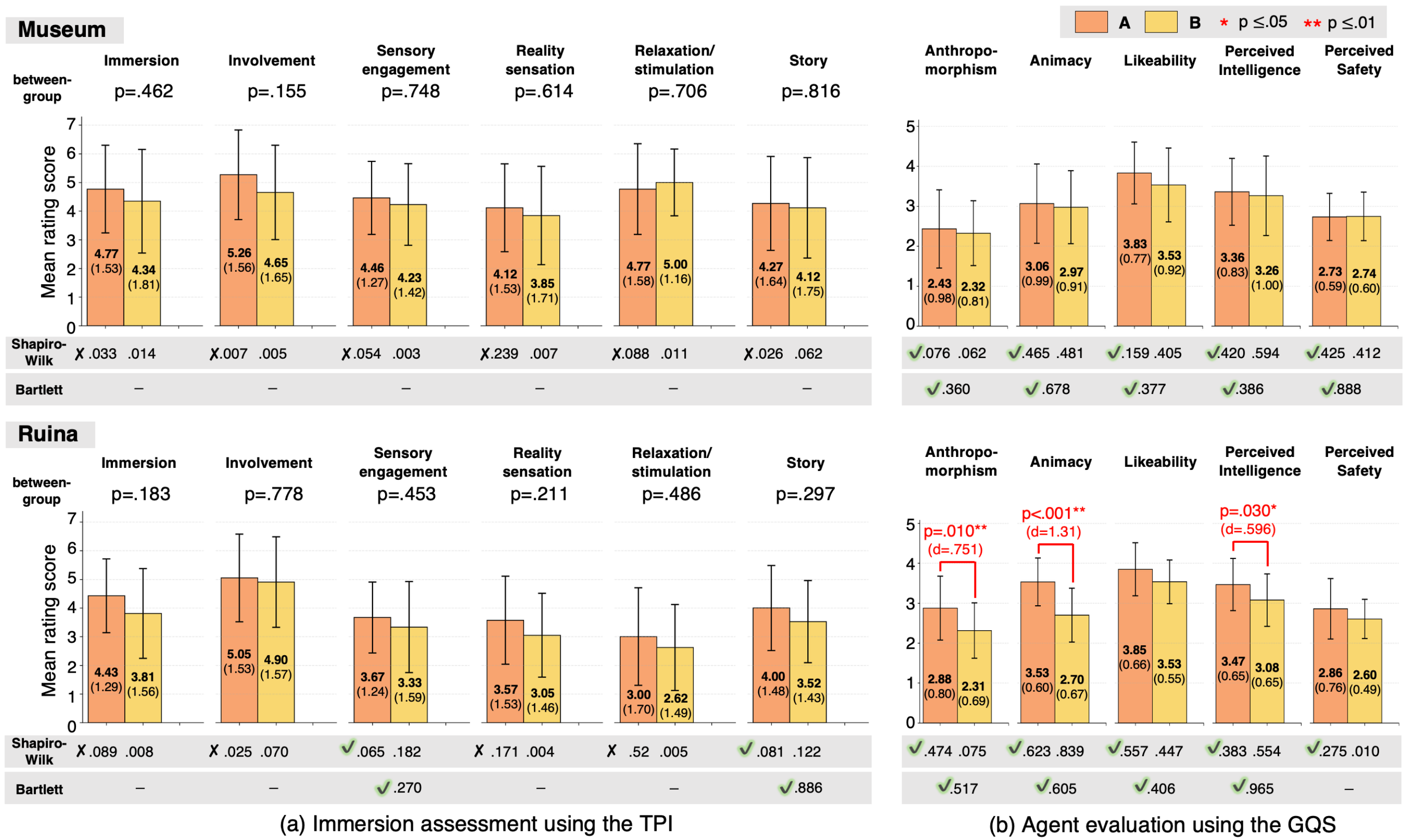}
    \vspace*{-3mm}
    \caption{Immersion experience and agent ratings across Museum and Ruina environments in VR-HMD study. Data presentation format and statistical analysis procedures follow those described in Fig.~\ref{fig:questionnaire_results_pc}. (a) TPI immersion assessment. (b) GQS agent evaluation.}
    \label{fig:questionnaire_results_hmd}
    \vspace{-3mm}
\end{figure*}

\paragraph{Immersion and Agent Evaluation} Fig.~\ref{fig:questionnaire_results_pc} and Fig.~\ref{fig:questionnaire_results_hmd} shows the results of immersion and agent evaluations. PC results showed Group A trending toward higher immersion scores in Museum without reaching statistical significance, while Ruina's "Relaxation/Stimulation" dimension indicated that agent presence enhanced environmental stimulation value. Group A consistently scored higher on "Animacy" and "Likeability" dimensions across both PC environments.

VR-HMD results showed environment-dependent effects: Museum showed no significant differences between Groups A and B across all dimensions, suggesting exhibition contexts may constrain differential agent effects in immersive settings. Conversely, Ruina demonstrated Group A superiority across three dimensions: "Anthropomorphism," "Animacy," and "Perceived Intelligence", indicating that VR-HMD platforms amplify social perception of on-demand agents specifically within social contexts.

\subsubsection{Experience Level Analysis}
We analyzed participant behavior based on prior platform usage, categorizing users with less than 50 hours of total platform time as novices and others as veterans. While sample sizes in some novice subgroups preclude robust statistical testing, mean exploration times consistently favored on-demand agents across experience levels. For example, representative VR-HMD results demonstrate this pattern: in Ruina, novices achieved 10.24~$\pm$~6.67 minutes of free exploration with Group A versus 1.44~$\pm$~1.12 minutes with Group B (n=6 vs. n=5), while veterans showed comparable advantages (10.03~$\pm$~9.62 vs. 1.23~$\pm$~0.90 minutes, n=15 vs. n=16). PC client and Museum data also revealed similar directional effects. These consistent patterns suggest Navigation Pixie benefits extend regardless of prior metaverse familiarity. Detailed statistical analyses are provided in Supplementary Material.

\subsubsection{Free-Text Responses} 
We categorized free-text responses from both PC client and VR-HMD studies and identified distinct thematic patterns for each experimental condition. Detailed comment examples are provided in Supplementary Material.

\paragraph{(A) On-demand Agent} Participants consistently provided positive feedback across four primary dimensions: \textbf{social presence}, \textbf{integration of speech understanding with spatial information}, \textbf{natural dialogue capabilities}, and \textbf{personalized experience}. VR-HMD users particularly emphasized enhanced character realism and social connection compared to PC users.
Technical limitations emerged consistently across platforms, with participants reporting \textbf{voice synthesis quality issues} and \textbf{spatial recognition inaccuracies}. 

\paragraph{(B) Fixed-route Agent} 
Participants acknowledged benefits of \textbf{structured guidance} and \textbf{multimodal explanations} while expressing significant dissatisfaction with \textbf{constrained exploration freedom} and \textbf{limited interactivity}. VR-HMD participants particularly emphasized the impersonal, mechanic nature of scripted responses within immersive environments.

\paragraph{(C) Control} PC client participants reported \textbf{spontaneous discovery} benefits but predominantly cited \textbf{absence of social interaction} and \textbf{insufficient contextual information}. 

\section{Discussion}
This study implemented and evaluated Navigation Pixie, an on-demand navigation agent, on the commercial metaverse platform Cluster, comparing it to fixed-route agent and no-agent conditions across both PC client and VR-HMD platforms. Results demonstrate that on-demand agents significantly increased dwell time and free exploration time across both platforms, though subjective evaluation patterns revealed distinct platform-specific and context-dependent effects.

This section discusses the analysis for each RQ, the differences between platforms, future improvement directions for on-demand navigation AI agents in commercial metaverse platforms, and current limitations of our study with directions for future research.

\subsection{RQ1: Enhancement of Exploration Behavior and Dwell Time}
The experimental results showed that Group A had 1.5-1.7 times longer dwell times and 3-5 times longer free exploration periods compared to Group B and Group C under both PC client and VR-HMD conditions. This is consistent with previous findings that users who can actively inquire, rather than passively receive explanations, exhibit more active behavior~\cite{8447553}.

Free-text responses and behavioral observations suggest that interactive exploration support with immediate responses increases curiosity and exploration motivation, not only in controlled VR settings, but also in UGC environments. VR-HMD participants expressed stronger emotional engagement with phrases like ``I wanted to see much more'' and ``it felt like exploring together,'' indicating heightened motivational effects in immersive contexts. In addition, the agent's detailed environmental descriptions extend the established finding that ``additional information provided by the guide increases exploration volume''~\cite{barneche2023assisted} to UGC contexts.

\subsection{RQ2: Interaction Format, Navigation Design, and User Experience}
The on-demand agent demonstrated the following key advantages and challenges with distinct manifestations across PC client and VR-HMD platforms:

\begin{enumerate} [leftmargin=*]
\setlength{\parskip}{0cm}
\item \textbf{Flexible Natural Language Dialogue and Personalized Experience}: Incorporating user-specific information into conversations was found to be critical to improving likeability. Referencing users' names, preferences, and past interactions is known to increase trust and attachment~\cite{10.1145/1067860.1067867}. The on-demand agent changed perceptions from a mere guide to a ``virtual partner'' by demonstrating interest and empathy toward users.

\item \textbf{Integration of Spatial Information with Detailed Explanations}: Group A users appreciated the high resolution of the spatial responses, indicating effective integration between the LLM's flexible responses and pre-configured navigation data. However, spatial recognition errors were also noted, highlighting the need for improved accuracy. 

\item \textbf{Immersion Reduction Due to Technical Limitations}: Free responses indicated that unnatural speech, slow response and hallucinations in GPT-4o real-time speech responses increased distrust of the agent. Previous research suggests that the naturalness of speech processing directly affects agent trustworthiness~\cite{10.1145/3706066, 10.3389/fpsyg.2025.1495456}, which may explain the lack of significant differences in the GQS dimensions of anthropomorphism, intelligence, and perceived safety. 
\end{enumerate}

\subsection{Embodiment Potential for On-Demand Navigation Agents}
We observed Group A users treating agents as human-like entities - saying ``see you later'' at the end of the experiment, or following agent accounts in anticipation of future encounters. These behaviors confirm that agent embodiment in VR spaces enhances social presence. This phenomenon is consistent with the ``Computers Are Social Actors'' paradigm~\cite{10.1111/0022-4537.00153}, where minimal social cues (voice or humanoid avatar) lead people to perceive agents as ``present companions'' and apply social rules to interactions~\cite{10.3389/frvir.2021.786665}.


Eliciting interactions similar to human-to-human exchanges requires a coherent integration of natural voice, facial expressions, and gestures that further enhance trust and familiarity~\cite{10.1145/3706066}. Previous research has suggested that NPCs can create a social atmosphere even in uninhabited spaces~\cite{10.3389/frvir.2023.1334795}, and our study clearly demonstrates that agents can provide users with ``community-like warmth'' beyond mere automated guidance.


\subsection{Platform-Specific Agent Perception and Environmental Context Dependencies}
PC environments showed consistent directional preferences for on-demand agents, regardless of the context. In contrast, VR-HMD environments exhibited strong environmental dependency. Exhibition contexts (museum) showed no differential agent effects, whereas social contexts (ruins) demonstrated the superiority of on-demand agents across the dimensions of anthropomorphism, animacy, and intelligence.

We attribute this phenomenon to context-specific evaluation mechanisms. In museum environments, users primarily evaluate agents as information delivery tools, and immersion in the experience reinforces functional assessment criteria. In contrast, VR-HMD immersion in Ruina environments enhances co-presence perception. This also occurs when fixed-route agents function appropriately in exhibition contexts but generate dissatisfaction in social spaces, where conversational flexibility is paramount. These platform-specific differences suggest that immersive devices function as cognitive amplifiers for agent presence perception.

\subsection{Approaches for Improving Agent Performance}
Several technical approaches could further improve the effectiveness of on-demand agents:

\begin{enumerate}[leftmargin=*]
\setlength{\parskip}{0cm}
\item \textbf{Dynamic Navigation Data Generation and Updates}: Integrating algorithms that automatically update NavMesh during exploration would allow flexible adaptation to newly added elements in changing UGC environments~\cite{10.1002/cav.1468}.

\item \textbf{Personalization Through Long-term Memory Modules}: Building on relational agent research~\cite{10.1145/1067860.1067867}, accumulating and referencing user-specific conversation histories would allow more intimate communication based on previous visit experiences.

\item \textbf{Integration of Multimodal Recognition and Non-verbal Expression}: Combining reinforcement learning-based agent~\cite{wang2023voyager} with visual and sensor data integration into LLMs could enablefiner-grained guidance through reduced object misidentification rates and real-time explanation of items users are focusing on~\cite{10.1145/1067860.1067867}.
\end{enumerate}

\subsection{Limitations and Future Directions}
Our research has several limitations that suggest directions for future work:

\paragraph{Temporal Constraints and Long-term Effects} 
This experiment measured effects within single sessions (approximately 30 minutes), leaving long-term behavioral changes and learning effects unexamined~\cite{10.7551/mitpress/8336.003.0011}. The observed ``following'' behavior toward agents suggests the need for longitudinal studies that examine ongoing relationship formation and its effects.

\paragraph{Participant Characteristics and Device Constraints} Our recruitment approach, which drew participants from the platform's existing user base, resulted in predominantly experienced users ($>$ 70\% weekly users). Preliminary behavioral metrics revealed directionally consistent agent effectiveness across experience levels, but limited novice sample sizes preclude definitive statistical conclusions. Furthermore, our evaluation excluded smartphone-based interactions.  Mobile platforms introduce distinct constraints through touch-based navigation and interface limitations that may fundamentally alter agent effectiveness patterns, which represent an important future investigation.

\paragraph{Influence of Diverse Design Elements}
While this study used a fairy-like avatar for the agent, the effect of different avatar designs (humanoid, animal, etc.) on user responses remains unexplored~\cite{10.1111/j.1468-2958.2007.00299.x}. Given the heightened anthropomorphic attribution observed in VR-HMD environments, systematic investigation of avatar design effects across platforms represents a particularly important research direction. Systematic comparative studies of avatar designs would help to establish optimal design guidelines for specific purposes.

\paragraph{Extending to Multi-Agent, Multi-User Environments} 
Real-world metaverse use involves the coexistence of multiple agents and users, necessitating research into navigation support in such environments. Agent designs that serve as social catalysts facilitating user interactions and supporting collaborative exploration are important directions for development.

\paragraph{User Co-Creation in Agent Development} 
Approaches where ``User A creates navigation data through dialogue that User B experiences'' demonstrate the potential of co-creative development using UGC. Investigating how such co-creation processes affect agent attachment and user satisfaction is an interesting research topic.

\section{Conclusion}
This paper presents Navigation Pixie, an LLM-based conversational navigation agent for commercial metaverse platforms. Our experiments in Cluster show that on-demand navigation agents significantly extend user engagement, increasing dwell time by 1.5-1.7 times and free exploration time by 3-5 times compared to fixed-route agents or no-guidance conditions.

Our key contributions include a loosely coupled architecture that separates platform-dependent components for improved portability, effective integration of LLM capabilities with spatial information to enable contextual guidance in UGC spaces, and a large-scale study demonstrating the effectiveness of on-demand agents in different virtual environments.
We found that the social presence of agents consistently increased exploration motivation; users naturally developed human-like interactions with embodied agents, consistent with the finding that minimal social cues trigger companion-like perceptions.

Our Navigation Pixie establishes a foundation for on-demand guidance frameworks that leverage the diversity of UGC. We anticipate that this work will establish new directions for metaverse agent design and advance human-centered interaction research in VR environments.


\acknowledgments{
This work was partially supported by JST Moonshot Research \& Development Program Grant Number JPMJMS2013 and JST ASPIRE Grant Number JPMJAP2327.

We would like to express our sincere gratitude to the creators who provided virtual worlds for our experiment: Taki, the creator of Ruina: A Small Cafe In The Sky, and Zawa, the creator of Museum.
}

\bibliographystyle{abbrv-doi}

\bibliography{template}

\balance

\ifarXiv
  

\section*{Supplementary material}

# Supplementary Material for
# Navigation Pixie: Implementation and Empirical Study Toward On-demand Navigation Agents in Commercial Metaverse

Hikari Yanagawa [*]
Cluster Metaverse Lab
University of Tsukuba

Yuichi Hiroi [†]
Cluster Metaverse Lab

Satomi Tokida [‡]
The University of Tokyo
Cluster Metaverse Lab

Yuji Hatada [§]
The University of Tokyo

Takefumi Hiraki [¶]
University of Tsukuba
Cluster Metaverse Lab

## 1 EXTENSIBILITY TO OTHER PLATFORMS

Our system achieves high extensibility by encapsulating platform-specific processing within the agent driver. Porting to other metaverse platforms (e.g., VRChat) requires only implementing the agent driver server for the target platform.

Two extension approaches are possible:

1. For platforms with API access: Implement the agent driver server using the platform's external control interfaces.
2. For platforms without API access: Implement a driver similar to Claude Computer Use[1] that externally controls the client, with UGC world information placed and referenced externally.

This architecture enables direct reuse of LLM-based dialog functionality and navigation strategies without modifications to Agent Core.

## 2 QUESTIONS RELATED TO IMMERSION RATING

For the immersion rating, we used the engagement dimension of the Temple Presence Inventory (TPI). The questions used are described below.

**Questionnaire**

Please tell us about your VR experience in this virtual world.

**To what extent did you feel mentally immersed in the experience?**

| Not at all | 1 | 2 | 3 | 4 | 5 | 6 | 7 | Very much |

**How involving was the experience?**

| Not at all | 1 | 2 | 3 | 4 | 5 | 6 | 7 | Very much |

**How completely were your senses engaged?**

| Not at all | 1 | 2 | 3 | 4 | 5 | 6 | 7 | Very much |

**To what extent did you experience a sensation of reality?**

| Not at all | 1 | 2 | 3 | 4 | 5 | 6 | 7 | Very much |

**How relaxing or stimulating was the experience?**

| Very relaxing | 1 | 2 | 3 | 4 | 5 | 6 | 7 | Very stimulating |

**How engaging was the story?**

[*] e-mail: h.yanagawa@cluster.mu
[†] e-mail: y.hiroi@cluster.mu
[‡] e-mail: tokidasatomi@g.ecc.u-tokyo.ac.jp
[§] e-mail: hatada@nae-lab.org
[¶] e-mail: hiraki@slis.tsukuba.ac.jp

[1] https://www.anthropic.com/news/3-5-models-and-computer-use

| Not at all | 1 | 2 | 3 | 4 | 5 | 6 | 7 | Very much |

## 3 QUESTIONS RELATED TO AGENT EVALUATION

For agent evaluation, we employed the Godspeed Questionnaire Series (GQS). The questions used are described below.

**Questionnaire**

Please evaluate your impressions of the navigation agent that guided you through the world in this experiment, based on the following scale.

### A. Anthropomorphism

| | | | | | | |
|---:|---|---|---|---|---|---|
| Fake | 1 | 2 | 3 | 4 | 5 | Natural |
| Machinelike | 1 | 2 | 3 | 4 | 5 | Humanlike |
| Unconscious | 1 | 2 | 3 | 4 | 5 | Conscious |
| Artificial | 1 | 2 | 3 | 4 | 5 | Lifelike |
| Rigidly | 1 | 2 | 3 | 4 | 5 | Elegantly |

### B. Animacy

| | | | | | | |
|---:|---|---|---|---|---|---|
| Dead | 1 | 2 | 3 | 4 | 5 | Alive |
| Stagnant | 1 | 2 | 3 | 4 | 5 | Lively |
| Mechanical | 1 | 2 | 3 | 4 | 5 | Organic |
| Artificial | 1 | 2 | 3 | 4 | 5 | Lifelike |
| Inert | 1 | 2 | 3 | 4 | 5 | Interactive |
| Apathetic | 1 | 2 | 3 | 4 | 5 | Responsive |

### C. Likeability

| | | | | | | |
|---:|---|---|---|---|---|---|
| Dislike | 1 | 2 | 3 | 4 | 5 | Like |
| Unfriendly | 1 | 2 | 3 | 4 | 5 | Friendly |
| Unkind | 1 | 2 | 3 | 4 | 5 | Kind |
| Unpleasant | 1 | 2 | 3 | 4 | 5 | Pleasant |
| Awful | 1 | 2 | 3 | 4 | 5 | Nice |

### D. Perceived Intelligence

| | | | | | | |
|---:|---|---|---|---|---|---|
| Incompetent | 1 | 2 | 3 | 4 | 5 | Competent |
| Ignorant | 1 | 2 | 3 | 4 | 5 | Knowledgeable |
| Irresponsible | 1 | 2 | 3 | 4 | 5 | Responsible |
| Unintelligent | 1 | 2 | 3 | 4 | 5 | Intelligent |
| Foolish | 1 | 2 | 3 | 4 | 5 | Sensible |

### E. Perceived Safety

| | | | | | | |
|---:|---|---|---|---|---|---|
| Anxious | 1 | 2 | 3 | 4 | 5 | Relaxed |
| Calm | 1 | 2 | 3 | 4 | 5 | Agitated |
| Quiescent | 1 | 2 | 3 | 4 | 5 | Surprised |

## 4 FREE RESPONSE ANSWERS

We included several free response questions in our study. The questions used are described below.

**Questionnaire**

- Did the agent-based guidance stimulate your motivation to explore the virtual world? If there were elements that stimulated your interest, which specific features were most effective?

- Please freely describe any other memorable impressions or suggestions for improvement from your overall experience.

## 5 EXPERIENCE LEVEL ANALYSIS

### 5.1 Methodology

We classified participants based on total platform usage time in the year preceding experimentation, establishing a 50-hour threshold to distinguish between novice ($< 50$ h) and veteran ($\geq 50$ h) users. This classification aimed to evaluate whether Navigation Pixie effectiveness varies with user familiarity with metaverse environments.

The experience level distribution revealed substantial imbalances across conditions. Novice sample sizes ranged from n=1 to n=6 per condition, while veteran groups contained n=10 to n=23 participants. These sample size constraints, particularly the single-participant groups in some conditions, preclude reliable statistical inference for novice populations. Consequently, we present descriptive statistics to characterize behavioral trends while acknowledging the exploratory nature of this analysis.

### 5.2 Results

**Dwell Time** Overall engagement patterns mirrored exploration behavior across experience levels. VR-HMD novices in Museum spent $15.98 \pm 6.93$ minutes (A, n=5) compared to $13.78 \pm 6.70$ minutes (B, n=3), while Ruina novices demonstrated $21.41 \pm 6.14$ minutes (A, n=6) versus $12.67 \pm 7.92$ minutes (B, n=5). Veteran users showed consistent on-demand agent advantages across both environments and platforms.

**Free Exploration Time** Novice users demonstrated substantial benefits from on-demand agents across both platforms. In VR-HMD environments, Museum novices showed 1.5-fold longer exploration with Navigation Pixie (A: $7.49 \pm 6.03$ min) compared to fixed-route guidance (B: $4.93 \pm 3.93$ min, n=3). The effect proved more pronounced in Ruina, where novice users with on-demand agents explored for $10.24 \pm 6.67$ minutes (n=6) versus $1.44 \pm 1.12$ minutes with fixed-route agents (n=5), representing a seven-fold difference.

PC client results exhibited similar patterns despite limited novice samples. Museum environments showed novice exploration times of $14.61 \pm 9.13$ minutes (A, n=4) versus $2.43 \pm 0.60$ minutes (B, n=2). Ruina novices demonstrated $3.60 \pm 2.75$ minutes of exploration with on-demand agents (n=5) compared to 0.58 minutes for the single fixed-route participant.

Veteran users consistently replicated these effects with larger, more stable samples. VR-HMD veterans in Museum environments explored for $10.08 \pm 5.55$ minutes (A, n=21) versus $3.07 \pm 2.29$ minutes (B, n=23). Ruina veterans showed nearly identical exploration duration (A: $10.03 \pm 9.62$ min, n=15) compared to minimal fixed-route exploration (B: $1.23 \pm 0.90$ min, n=16).

## 6 DETAILED FREE-TEXT RESPONSE EXAMPLES

This section provides detailed participant feedback across both PC client and VR-HMD experimental conditions.

### 6.1 Integrated User Response Analysis

(A) On-demand Agent **Social presence and natural dialogue** emerged as primary positive themes, with users valuing the sense of companionship and responsive interaction across both platforms:

- *"I enjoyed having someone beside me, exploring the world together."* (PC)

- *"The agent responded to my questions, which made me want to keep talking to it."* (PC)

- *"When I asked about notable features of the world, the agent responded with details ... the high resolution of these responses was intriguing."* (PC)

- *"I felt a sense of human-like quality in how the navigation agent reasoned about the exhibition sequence, which motivated me to wonder where it would guide me next."* (VR)

- *"The character had a sense of presence and matched the world's atmosphere."* (VR)

- *"I felt like there was a staff member present rather than just a decoration."* (VR)

- *"Being able to communicate with the agent made me feel like I was having a conversation."* (VR)

**Personalized experience** features generated particularly positive responses, demonstrating the effectiveness of user-specific information integration:

- *"I was pleased when it called me by my nickname after reading my profile."* (PC)

- *"I was happy when it called my name at the end, which enhanced immersion."* (VR)

**Enhanced spatial understanding** emerged as a distinct advantage in VR environments:

- *"Seeing dinosaur bones in 3D through VR helped me understand their structure well."* (VR)

- *"The museum concept with navigator guidance was excellent."* (VR)

- *"The fairy-like appearance and movement, along with how it spoke to me."* (VR)

However, **technical limitations** were consistently identified across both platforms, particularly regarding content accuracy and system reliability:

- *"Sometimes it would start reading text that wasn't in the comments, resulting in distorted content."* (PC)

- *"It insisted there was a second floor when there wasn't."* (PC)

- *"Voice playback was cut off several times."* (VR)

- *"There were audio bugs that prevented me from feeling much immersion despite entering in VR."* (VR)

(B) Fixed-route Agent **Structured guidance benefits** were recognized for comprehensive coverage:

- *"I felt reassured that the tour was planned to cover everything without missing anything."* (PC)

- *"Having explanations in both voice and text was helpful."* (PC)

Conversely, **exploration constraints** generated significant dissatisfaction, with participants emphasizing the restrictive nature of predetermined routes:

- *"Being limited to a fixed route felt restrictive because I couldn't freely explore areas of interest."* (PC)

- *"Since it didn't react, but only provided guidance, I didn't feel there was anything particularly new about the experience."* (PC)

- *"My impression is that it did not inspire exploration. Agents inorganic and mechanically set in motion."* (VR)

**(C) Control** Control condition participants demonstrated mixed responses, with limited positive discovery experiences overshadowed by social isolation concerns. **Spontaneous discovery** provided some engagement:

– *"It was enjoyable when I discovered areas with interactive elements."* (PC)

However, **social isolation and information deficiency** dominated feedback:

– *"It felt lonely with no one else in the cafe."* (PC)

– *"I wanted to tell friends about this or take photos together to share with others."* (PC)

– *"There was a lack of background information, which made some things difficult to understand."* (PC)


\fi

\end{document}